\documentclass[journal=jpcl,manuscript=article]{achemso}
\usepackage{bm}
\usepackage{caption}
\usepackage{subcaption}
\usepackage{graphicx}
\usepackage{float}
\usepackage{array}
\usepackage{multirow}
\usepackage{xcolor}
\usepackage{color}
\usepackage[hidelinks,breaklinks,colorlinks=true,linkcolor=blue,citecolor=blue,urlcolor=blue]{hyperref}
\usepackage[version=3]{mhchem} 

\author{Nikhil R. Agrawal}
\affiliation[University of California, Berkeley]
{Department of Chemical and Biomolecular Engineering, University of California, Berkeley, CA 94720-1462, USA}
\author{Chao Duan}
\affiliation[University of California, Berkeley]
{Department of Chemical and Biomolecular Engineering, University of California, Berkeley, CA 94720-1462, USA}
\author{Rui Wang}
\email{ruiwang325@berkeley.edu}
\affiliation[University of California, Berkeley]
{Department of Chemical and Biomolecular Engineering, University of California, Berkeley, CA 94720-1462, USA}
\alsoaffiliation[Lawrence Berkeley National Laboratory]
{Materials Sciences Division, Lawrence Berkeley National Laboratory, Berkeley, 94720, California, USA}

\title{On the nature of overcharging and charge inversion in electrical double layers}

\abbreviations{IR,NMR,UV}

\begin{document}

\begin{tocentry}

\begin{figure}[H]
\captionsetup[subfigure]{labelformat=empty}
 \includegraphics[width=1\columnwidth]{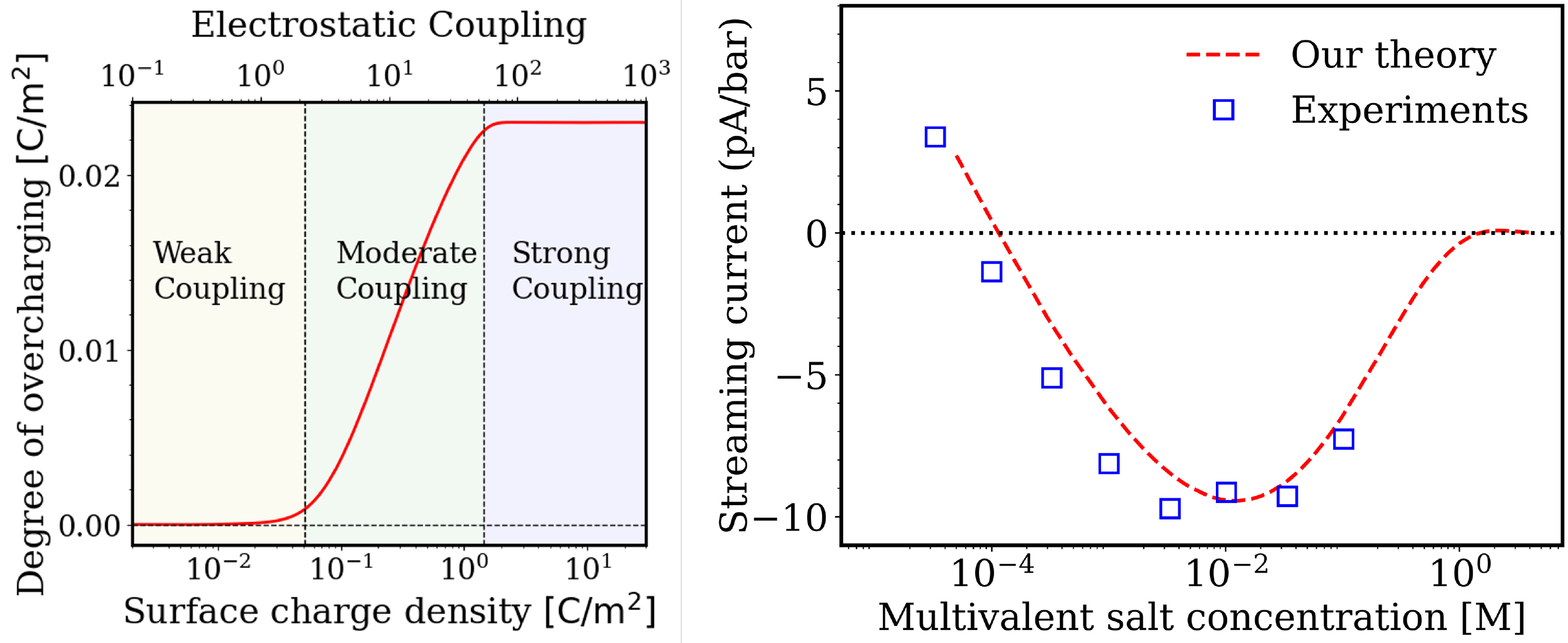}
    \caption*{}
\end{figure}

\end{tocentry}

\begin{abstract}
Understanding overcharging and charge inversion is one of the long-standing challenges
in soft matter and biophysics. To study these phenomena, we employ the modified Gaussian renormalized fluctuation theory, which allows for the self-consistent accounting of spatially varying ionic strength, as well as the spatial variations in dielectric permittivity and excluded volume effects. The underlying dependence of overcharging on the electrostatic coupling is elucidated by varying surface charge, counterion valency, and dielectric contrast. Consistent with simulations, three characteristic regimes corresponding to weak, moderate, and strong coupling are identified. Important features like the inversion of zeta potential, crowding and ionic layering at the surface are successfully captured. For weak coupling, there is no overcharging. In the moderate coupling regime, overcharging increases with surface charge. Finally, in the strong coupling regime, ionic crowding and saturation in overcharging are observed. Our theory predicts non-monotonic dependence of charge inversion on multivalent salt concentration as well as the addition of monovalent salt, in quantitative agreement with experiments.
\end{abstract}
\pagebreak
\section{Introduction}
Modeling electrical double layers (EDL) is of vital importance to the field of soft matter physics. Although the classical mean-field Poisson-Boltzmann (PB) theory is physically intuitive and numerically soluble\cite{Lu2008RecentApplications}, it does not account for three essential factors: ion correlations, dielectric variation, and excluded volumes of ions and solvent. Capturing these missing factors is needed for the fundamental understanding of many phenomena crucial to protein stability\cite{Grosberg2002Colloquium:Systems, Zhang2008ReentrantCounterions}, aerosols in atmospheric chemistry\cite{Knipping2000ExperimentsAerosols}, energy storage devices\cite{JohnNewman2012ElectrochemicalEdition, Sing2014ElectrostaticMorphology,Fedorov2014IonicInterfaces, He2009TuningInversion,Gillespie2012HighChannels}, biomedical materials\cite{Gelbart2007DNAInspiredElectrostatics, Felgner1997NonviralTherapy}, and numerous other physicochemical and biophysical systems \cite{Krishnamoorthy2014Surface-initiatedCoatings,Besteman2007ChargeIons,Tata2006ColloidalColloids,Yu2018MultivalentBrushes, Lobaskin2007ElectrophoresisRegime}.\par 

One long-standing puzzle beyond the scope of PB is the over-accumulation of counterions near a charged surface, known as overcharging \cite{Besteman2004DirectPhenomenon, Grosberg2002Colloquium:Systems, Kubickova2012OverchargingCations}. The overcharging of EDL may lead to a reversal in the sign of electrophoretic mobility of colloidal particles or in the direction of ionic current in electro-osmotic flows. This reversal in the sign of zeta potential ($\psi_\mathrm{\zeta}$) is usually known as charge inversion \cite{Semenov2013ElectrophoreticSimulations, Lin2020ChargeNanopores, Martin-Molina2008ChargeSimulations, VanDerHeyden2006ChargeCurrents,Buyukdagliinversion}. For monovalent salts, it is well-accepted that the electrostatic correlation alone is not sufficient
to cause charge inversion in aqueous solutions \cite{vernin11inversion}. On the contrary, in the case of multivalent salts, charge inversion is overwhelmingly driven by ion correlations\cite{Besteman2004DirectPhenomenon,Kubickova2012OverchargingCations}. Experiments and simulations show a continuous transition from a normal double layer to an overcharged double layer as surface charge increases\cite{Diehl2006SmoluchowskiReversal, Diehl2008ColloidalConcentration, Martin-Molina2009EffectReversal}. Continuously increasing surface charge slows down overcharging and eventually leads to ionic crowding at the surface\cite{Diehl2006SmoluchowskiReversal, Kornyshev2007Double-layerChange,Bazant2011DoubleCrowding}. The effect of salt concentration on charge inversion is also non-trivial and shows a non-monotonic change in the magnitude of the inverted mobility and ionic current. \cite{VanDerHeyden2006ChargeCurrents, Martin-Molina2008ChargeSimulations, Hsiao2006Salt-inducedPolyelectrolytes}. Non-monotonic decrease in $\psi_\mathrm{\zeta}$ has also been observed when monovalent salt is added to a multivalent salt solution\cite{VanDerHeyden2006ChargeCurrents}. Furthermore, simulations show ionic layering and oscillation of electrostatic potential in the strong-coupling condition\cite{Hsiao2008OverchargingStudy,Kubickova2012OverchargingCations,Mezger2008MolecularSurface,DeSouza2020InterfacialLiquids}.   \par 

Many theoretical efforts have been made to model ion correlations and hence charge inversion\cite{Grosberg2002Colloquium:Systems,Netz2003VariationalSystems,Pianegonda2005ChargeParticles,mashayakaluru2018,santoslevinpair2010,Buyukdagli_2016,Gupta2020IonicModel,Quesada-Perez2003OverchargingApproach,Lau2008FluctuationSolution,Bazant2011DoubleCrowding,Gillespie2011EfficientlyTheory, hoff_gillespie2013,valiskosim_2018,lue2006variational,lue2009field,palmeri2010,palmericapillary2010}. The Strongly Correlated Liquid (SCL) theory\cite{Rouzina1996MacroionCloud, Perel1999ScreeningInversion, Shklovskii1999ScreeningCharge,Nguyen2000MacroionsCharge} assumes a two-dimensional condensed layer of counterions in equilibrium with a diffuse double layer described by mean-field PB. The presumption of a condensed layer excludes the possibility of capturing the transition from a normal diffuse double layer to an overcharged one. Lau \cite{Lau2008FluctuationSolution} developed a perturbative theory with a one-loop correction of the electrostatic potential. Their point-charge model overestimates the correlation and does not account for the excluded volume effect. Bazant et al.\cite{Bazant2011DoubleCrowding, Storey2012EffectsPhenomena, DeSouza2020InterfacialLiquids}, and Gupta et al.\cite{Gupta2020IonicModel} express the correlation in terms of different powers of electrostatic potential gradient which vanishes in the bulk thus ignoring the ion correlations there. Integral-equation-based approaches with hypernetted chain and mean-spherical approximation closures have also been used\cite{Martin-Molina2003LookingElectrolytes, jimenezcassou2004, jimenezcassou2008,teranscrivenpart1,teranscrivenpart2,waismanocharge1972,xinwuinversion}. All the above theories cannot fully capture the nonmonotonic dependence of charge inversion on salt concentration\cite{Storey2012EffectsPhenomena,Martin-Molina2008ChargeSimulations,stout_khair_2014}. Gillespie et al.\cite{Gillespie2011EfficientlyTheory,hoff_gillespie2013,voukadinova_gillespiedft} developed a density functional theory (DFT) that is able to reproduce the non-monotonic behavior. However, the reference ion concentration profile chosen for perturbation depends on density weighting functions that are specific to a particular system, preventing its generalization to other correlation-induced phenomena\cite{rosenfelddft1997, wulidft2007, Gillespie2002CouplingFlux}. Furthermore, DFT calculations are also computationally challenging.  \par 

Overcharging and charge inversion are relevant to the design of modern nanodevices and various biophysical processes. Therefore, it is desirable to develop a self-consistent and numerically solvable approach to model these phenomena. Localized charge inversion has proven to be crucial in the functioning of nanofluidic devices such as ionic diodes and rectifiers\cite{He2009TuningInversion, Lin2020ChargeNanopores}. The reversal of electrophoretic mobility offers a promising avenue for the development of innovative DNA sequencing methodologies\cite{Luan_2010,buyukdagli_dna_motion}. Furthermore, overcharging plays a vital role in the formation of chromatin; for instance, the amount of DNA wrapping around positive histone proteins in a nucleosome exceeds the requirements of charge neutrality by a significant margin\cite{chromatinphysics}. The most challenging task in modeling overcharged EDLs is to capture the inhomogeneous ion correlations as a result of spatially varying ion density from the surface to the bulk. This inhomogeneity is particularly significant for the case of charge inversion where the correlation is substantially stronger near the surface. In our previous work\cite{agrawalciwkb}, we employed a combination of WKB-like approximation and a boundary layer approach to separately model the correlation near the surface and in the diffused double layer. However, this treatment ignores the long-range feature of ion correlation which then overestimates the inverted $\psi_\mathrm{\zeta}$ and incorrectly predicts a discontinuous transition with surface charge. It also fails to capture correlation-induced oscillations in the ion density profiles. In this work, we apply the modified Gaussian renormalized fluctuation theory to study EDLs next to planar surfaces and account for both short-range and long-range features of ion correlation. This theory self-consistently includes spatially varying correlations, image charge effect, and excluded volumes in a unified framework. The nature of overcharging and charge inversion with respect to surface charge, counterion valency, salt concentration, the addition of monovalent counterions, and dielectric contrast is revealed. The predictions of our theory are in good agreement with experiments and simulation results. \par 

\section{Theory}
We consider a charged plate at $z=0$ with uniform surface charge density $\sigma$ in contact with an electrolyte solution containing cations of valency $q_+$ and anions of valency $q_-$. The dielectric function of the medium is given by $\varepsilon (z)$ and the excluded volumes of molecules by $v_\mathrm{\pm,s}$. To avoid the overestimation of charge interactions aroused by the point-charge model, we consider a finite spread of ionic charge given by distribution function $h_\pm(\mathbf{r}-\mathbf{r'})$. We will use the modified Gaussian renormalized fluctuation theory derived in our previous work\cite{agrawalciwkb} to model EDLs for this system. Compared to the earlier work from Z.-G. Wang and coworkers\cite{Wang2010FluctuationEnergy,Wang2013EffectsForces,Wang2015OnSurfaces}, there are two major modifications incorporated in our theory. First, the excluded volume effect of the molecules is systematically included in the grand canonical partition function to avoid the overaccumulation of ions at the surface. Second, the ion correlation is decoupled into a short-range contribution associated with the local electrostatic environment and a long-range contribution accounting for the spatially varying ionic strength and dielectric permittivity. This modified theory yields the following set of self-consistent equations for the non-dimensionalized electrostatic potential $\psi$, ion concentration $c_\pm$, self-energy of ions $u_\pm$, and correlation function $G$ 
 \begin{eqnarray}
{-\nabla.[\epsilon(z)\nabla\psi(z)]}= \sigma\delta(z) + q_{+}{c_{+}(z)} - q_{-}{c_{-}(z)} 
\label{eq:psi}
\end{eqnarray}
\begin{eqnarray}
{c_{\pm}(z)}= \frac{ \mathrm{e}^{\mu_{\pm}}}{v_{\pm}}\exp[\mp q_{\pm}\psi(z) - u_{\pm}(z) -v_{\pm} \eta(z)]
\label{eq:conc}
\end{eqnarray}
\begin{eqnarray}
{\textit{u}_{\pm}(\mathbf{r})}=\frac{q_{\pm}^2}{2}\int d\mathbf{r}'d\mathbf{r}'^{\prime}h_{\pm}(\mathbf{r'},\mathbf{r})G(\mathbf{r'},\mathbf{r'^{\prime}})h_{\pm}(\mathbf{r'^{\prime}},\mathbf{r})
\label{eq:selfe}
\end{eqnarray} 
\begin{eqnarray}
{-\nabla_{\mathbf{r'}}.[\epsilon(\mathbf{r'})\nabla_{\mathbf{r'}}G(\mathbf{r'},\mathbf{r'^{\prime}})]} + 2I(\mathbf{r'})G(\mathbf{r'},\mathbf{r'^{\prime}}) = \delta(\mathbf{r'}-\mathbf{r'^{\prime}})
\label{eq:greens}
\end{eqnarray}
where $\epsilon({\mathbf{r}})=kT\varepsilon_{0}\varepsilon ({\mathbf r})/e^2$ is the scaled permittivity with $\varepsilon_{0}$ as the vacuum permittivity and $e$ as the elementary charge. $\mu_{\pm}$ are chemical potentials of ions determined from the bulk salt concentration $c_\mathrm{b}$. $2I(\mathbf{r})= \epsilon(\mathbf{r})\kappa^2(\mathbf{r}) = c_{+}(\mathbf{r})q_{+}^2 + c_{-}(\mathbf{r})q_{-}^2$, with $I(\mathbf{r})$ and $\kappa(\mathbf{r})$ the local ionic strength and the inverse of the screening length, respectively. $\eta(\mathbf{r})$ is the field accounting for the excluded volume effect and is given by
\begin{equation}
\eta(z) =  -\frac{1}{v_s}{\ln[1 - v_+c_+(z) -v_- c_-(z)]}
\label{eq:eta}
\end{equation}\par
In the homogeneous bulk reservoir, $c_\pm(\mathbf{r})$ and $\epsilon(\mathbf{r})$ are constant and $\psi$ can be set to zero. Thus, using Eq. \ref{eq:conc} the chemical potential can be written as
\begin{equation}
\mu_\mathrm{\pm}= u_\mathrm{\pm,b} + \ln{c_\mathrm{\pm,b}v_\mathrm{\pm}}  +v_\mathrm{\pm}\eta_\mathrm{b}
\label{eq:mu_ions}
\end{equation}
where the subscript $b$ stands for bulk. For correlation function $G_\mathrm{b}$ in the bulk, Eq. \ref{eq:greens} has an analytical solution given by
\begin{equation}
G_\mathrm{b}(\mathbf{r'},\mathbf{r''})  =  \frac{\mathrm{e}^{-\kappa_\mathrm{b}|\mathbf{r'}-\mathbf{r''}|}}{{4\pi\epsilon_\mathrm{b}}|\mathbf{r'}-\mathbf{r''}|}
\label{eq:gbulk}
\end{equation} 
Next, we use the following mathematically convenient Gaussian form for charge distribution function $h_\pm(\mathbf{r}-\mathbf{r'})$ 
\begin{equation}
{h_{\pm}(\mathbf{r} -\mathbf{r'})}  =  \left({\frac{1}{2a_{\pm}}}\right)^{3/2}\exp\left[\frac{-\pi{}(\mathbf{r}-\mathbf{r'})^{2}}{2a^2_\pm}\right]
\label{eq:cspread}
\end{equation} 
where $a_\pm$ is the Born-radius of the ions. Combining Eq. \ref{eq:selfe} and Eq. \ref {eq:cspread} the following analytical expression for $u_\mathrm{\pm,b}$ can be derived 
\begin{equation}
u_\mathrm{\pm,b} = \frac{q^2_\pm}{8\pi\epsilon_\mathrm{b}} \left[\frac{1}{a_\mathrm{\pm}}-  \kappa_\mathrm{b}\exp\left(\frac{({a_\pm\kappa_\mathrm{b}})^2}{\pi}\right)\operatorname{erfc}\left(\frac{{a_\pm\kappa_\mathrm{b}}}{\sqrt{\pi}}\right)\right]
\label{eq:selfe_bulk}
\end{equation}
where the first term is the Born solvation energy and the second term is the contribution from the ion
correlation. Through $u_\mathrm{\pm,b}$ and $\eta_\mathrm{b}$ the chemical potential $\mu_\pm$ depends on the size of ions.\par

To accurately calculate the self-energy in the interface, we need to resolve the correlation function at two very different length scales, i.e., the length scale of the double layer as well as that of the ion size. We here use a decomposition scheme developed in our previous work \cite{agrawalvapliq} to solve this dual-length scale problem in a numerically tractable manner. The total correlation function is decoupled into the short-range correlation associated with ion size and long-range correlation associated with screening length as
\begin{eqnarray}
G(\mathbf{r'},\mathbf{r''}) = G_\mathrm{s}(\mathbf{r'},\mathbf{r''}) + G_\mathrm{l}(\mathbf{r'},\mathbf{r''})
\label{eq:greens_dec}
\end{eqnarray}
The short-range component $G_\mathrm{s}$ accounts for the local electrostatic environment whereas the long-range component $G_\mathrm{l}$ takes care of spatially varying ionic strength and dielectric permittivity. The length scale of double layers is usually a few nanometers, while the length scale of the ion is around 1-4 \AA. The width of the double layer is one order of magnitude larger than the ion size. Therefore, in most cases, the change in ionic strength is very small within the distance of the ion size. Hence, we construct the short-range correlation function $G_s$ using the local ionic strength $I(\mathbf{r})$ and dielectric permittivity $\epsilon(\mathbf{r})$ as follows:
\begin{equation}
-\epsilon(\mathbf{r}){\nabla_{\mathbf{r'}}^2G_\mathrm{s}(\mathbf{r'},\mathbf{r'^{\prime}})} + 2I(\mathbf{r})G_\mathrm{s}(\mathbf{r'},\mathbf{r'^{\prime}}) = \delta(\mathbf{r'},\mathbf{r'^{\prime}})
\label{eq:greens_short}
\end{equation}
The above equation has the same form as Eq. \ref{eq:greens}, however, the spatially varying ionic strength and dielectric permittivity are replaced by their local counterparts. $G_\mathrm{s}$ has a Debye-H\"{u}ckel style analytical form,
\begin{eqnarray}
G_\mathrm{s}(\mathbf{r'},\mathbf{r'^{\prime}})  =  \frac{\mathrm{e}^{-\kappa(\mathbf{r})|\mathbf{r'}-\mathbf{r'^{\prime}}|}}{4\pi\epsilon(\mathbf{r})|\mathbf{r'}-\mathbf{r'^{\prime}}|}
\label{eq:gsol_short}
\end{eqnarray} 
and $G_\mathrm{l}$ is given by
\begin{eqnarray}
{-\nabla_{\mathbf{r'}}.[\epsilon(\mathbf{r'})\nabla_{\mathbf{r'}}G_\mathrm{l}(\mathbf{r'},\mathbf{r'^{\prime}})]} + 2I(\mathbf{r'})G_\mathrm{l}(\mathbf{r'},\mathbf{r'^{\prime}})= S(\mathbf{r'},\mathbf{r'^{\prime}})
\label{eq:greens_long}
\end{eqnarray}
where the non-local source term $S$ is
\begin{equation}
S(\mathbf{r'},\mathbf{r'^{\prime}})   = \nabla_{\mathbf{r'}}.((\epsilon(\mathbf{r'})-\epsilon(\mathbf{r}))\nabla_{\mathbf{r'}}G_\mathrm{s}(\mathbf{r'},\mathbf{r'^{\prime}})) - 2(I(\mathbf{r'}) - I(\mathbf{r}))G_\mathrm{s}(\mathbf{r'},\mathbf{r'^{\prime}}) 
\end{equation}
The resulting self-energy $u_\mathrm{\pm}$ is 
\begin{eqnarray}
\textit{u}_{\pm}(z)=  \frac{q_{\pm}^2}{2}\int _{\mathbf{r}',\mathbf{r}'^{\prime}}h_{\pm}G_\mathrm{s}h_{\pm} + \frac{q_{\pm}^2}{2}G_\mathrm{l}(z,z)
\label{eq:selfe_hybrid}
\end{eqnarray} 

The first term on the r.h.s of the above equation is the short-range component of the self-energy $u_{\pm,s}$ given by
\begin{equation}
u_\mathrm{\pm,s}(z) =  \frac{q^2_\pm}{8\pi\epsilon(z)a_\mathrm{\pm}} -\frac{q^2_\pm \kappa(z)}{8\pi\epsilon(z)}\exp\left(\frac{{a_\pm^2\kappa(z)}^2}{\pi}\right)\mathrm{erfc}\left(\frac{{a_\pm\kappa(z)}}{\sqrt{\pi}}\right)
\label{eq:selfe_short}
\end{equation}
The second term of Eq. \ref{eq:selfe_hybrid} corresponds to the long-range contribution from $G_\mathrm{l}$ which is evaluated in the point-charge limit. Since the charge spread on the ion will be crucial to only the electrostatic forces originating in the close neighborhood of the test ion, the charge distribution function $h$ is retained while calculating $u_{\pm,s}(z)$. On the other hand, any significant change in ionic strength or dielectric permittivity that occurs over the length scale of the interface is attributed to $G_{l}$. Eq. \ref{eq:greens_long} shows that $G_{l}$ captures all the electrostatic effects that act at a much larger length scale than the ion size. Since these long-range effects will be oblivious to the shape of the ion, the
charge distribution functions $h$ can be replaced with a point charge model to facilitate the calculation of the self-energy. In the extreme case where the double layer thickness is comparable to the ion size, we can directly discretize the correlation function using a single grid size and the above decomposition procedure is not necessary. However, for simplicity, all the results shown in this paper are obtained using this decomposition scheme. The detailed derivation of the theory and the numerical scheme is provided in the Supplemental Material \cite{Xu2014SolvingEquations,suppinfo}.\par

\section{Results and Discussion}
In the current work, we study overcharging and charge inversion in the case of a negatively charged surface in contact with an aqueous electrolyte solution. Although, the equations above can account for local dielectric variations, for simplicity we solve for the case of the primitive model of electrolytes. $\varepsilon(z)$ is taken to be a step function with the value $\varepsilon_\mathrm{P}$ for $z<0$ and $\varepsilon_\mathrm{S} = 80$ for $z>0$. The salt solution is confined to the region $z>0$. In this article, the focus is on the effect of surface charge density, counterion valency, salt concentration, and dielectric contrast. Therefore, $q_-$ is set to 1, and ions and solvent molecules are considered to have the same radius $a$. Furthermore, we write excluded volumes as $v_\pm = \frac{4}{3}\pi a^3$, thus excluding the effect of the hydration shell of ions on the EDL structure.\par

\begin{figure*}
\captionsetup[subfigure]{labelformat=empty}
    \begin{subfigure}{0.32\columnwidth}
    \includegraphics[width=\columnwidth]{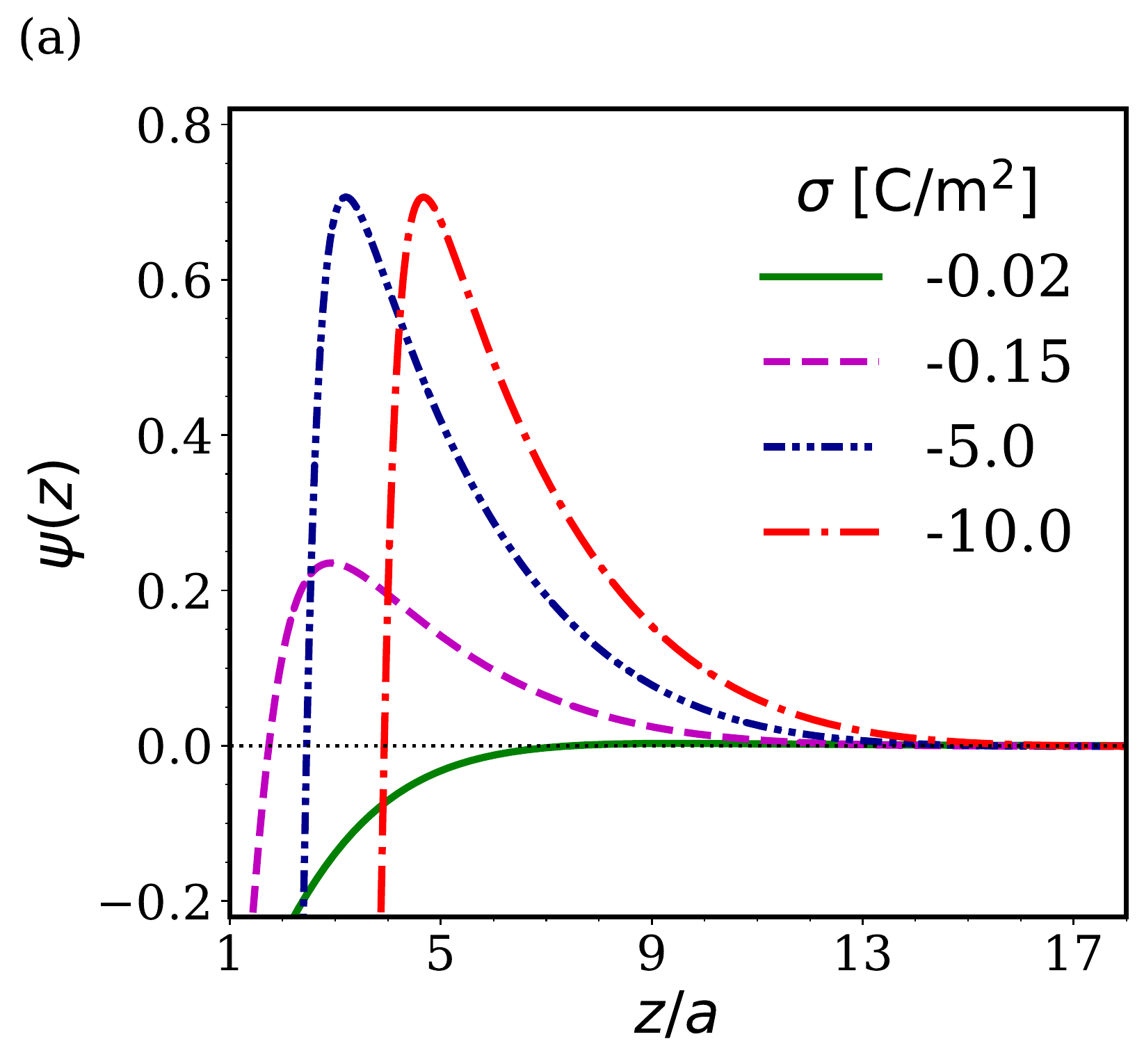}
    \caption{}
    \label{fig:psi_sigma}
    \end{subfigure}  
    \hfill
    \begin{subfigure}{0.32\columnwidth}
        \includegraphics[width=\columnwidth]{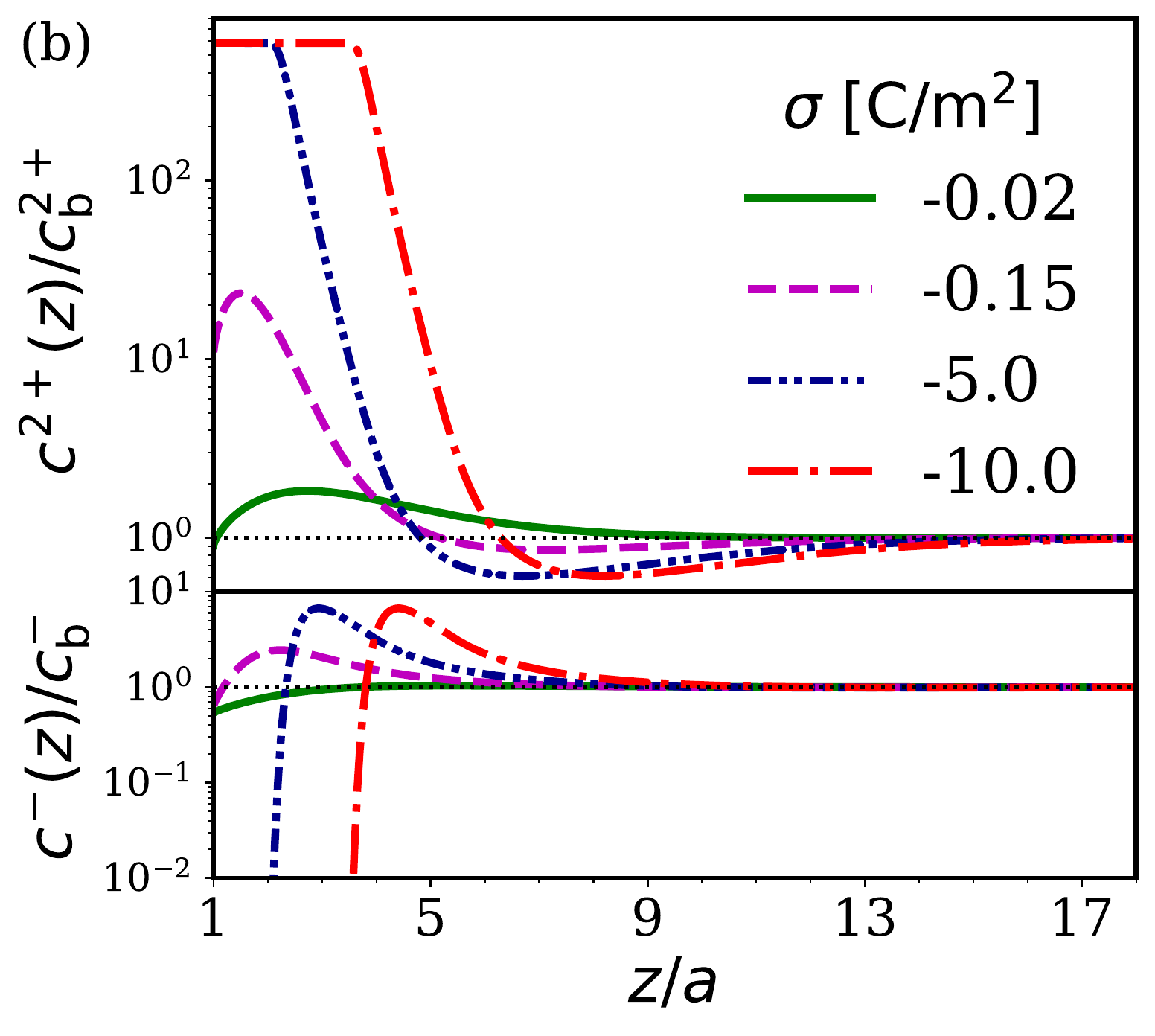}
        \caption{}
        \label{fig:conc_sigma}
    \end{subfigure} 
    \hfill
        \begin{subfigure}{0.32\columnwidth}
        \includegraphics[width=\columnwidth]{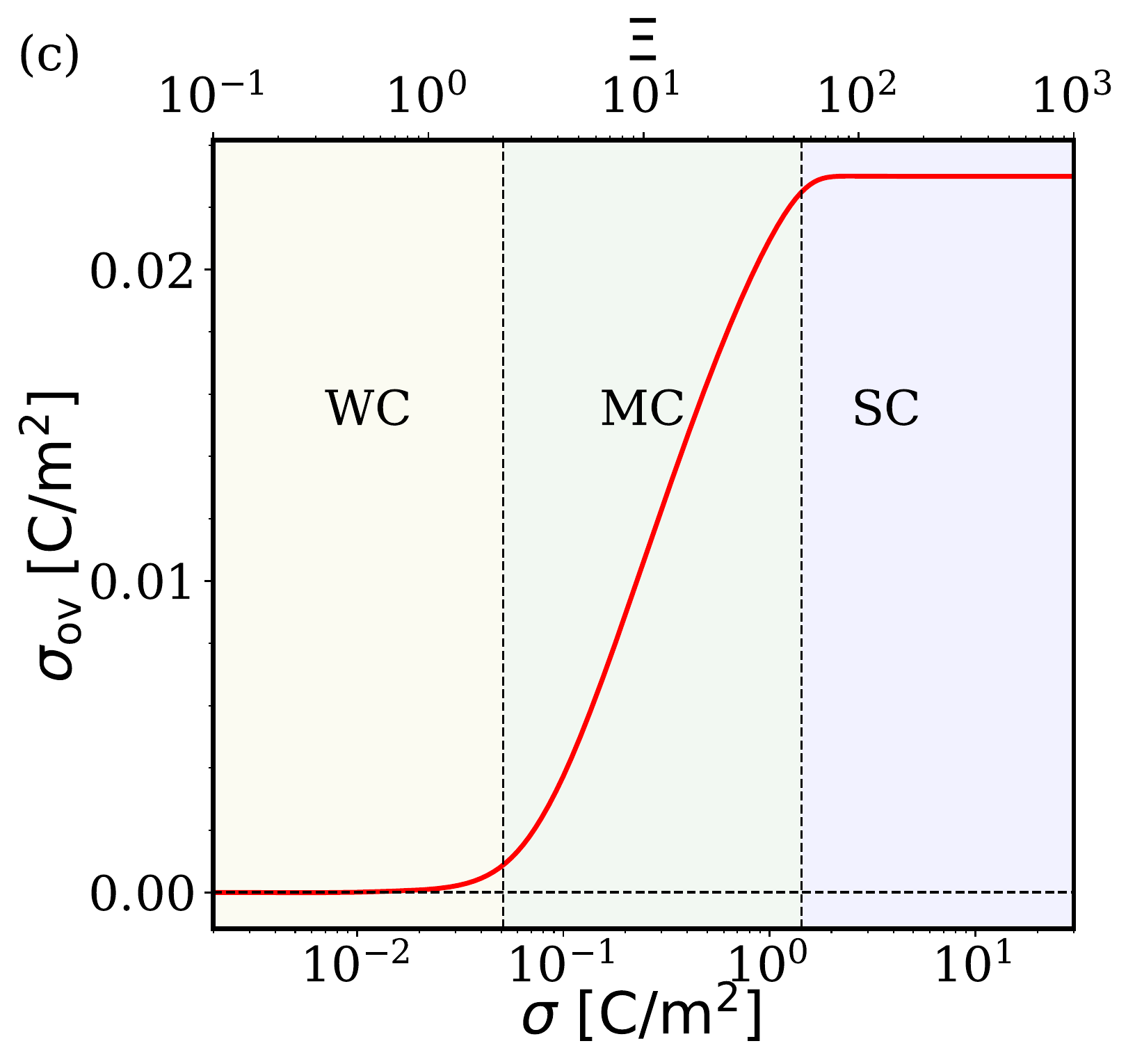}
        \caption{}
        \label{fig:ov_sigma}
    \end{subfigure} 
\caption{Continuous transition from normal double to overcharged double layer depicted using 2:1 salt solution. $c_\mathrm{b}$ = 0.2 M, $a_{\pm,s} = 1.5$ \AA\ and $\varepsilon_\mathrm{S} = \varepsilon_\mathrm{P} = 80$. a) Electrostatic potential $\psi(z)$ profiles and b) distributions of counterion and coions for increasing surface charge density $\sigma$. c) The degree of overcharging $\sigma_\mathrm{ov}$ as function of $\sigma$ and electrostatic coupling parameter $\Xi$. Three regions correspond to weak coupling (WC), moderate coupling (MC), and strong coupling (SC). }
\label{fig:panel}
\end{figure*}
Our theory successfully captures the transition from a normal double layer to an overcharged one as surface charge density increases. Fig. \ref{fig:psi_sigma} and \ref{fig:conc_sigma} show the electrostatic potential profile and ion distribution respectively for the case of a 2:1 salt solution. At a low $\sigma$ value of $-0.02$ C/m$^2$, the potential is negative in the entire region, in line with the normal double-layer structure predicted by PB. As $\sigma$ increases to $-0.15$ C/m$^2$, more counterions are attracted to the surface, enhancing the strength of ion correlations. Compared to the case of $\sigma = -0.02$ C/m$^2$, the EDL becomes narrower and the ion concentrations close to the surface increase for $\sigma = -0.15$ C/m$^2$, as has been observed in simulations \cite{valiskosim_2018, gillespie2015review}. 
The sign of potential turns from negative to positive, leading to an overcharged double layer. Because of overcharging, coions are enriched in the diffuse region far away from the surface, and counterions are depleted as depicted in Fig. \ref{fig:conc_sigma}. For very high $|\sigma| > 4$ C/m$^2$, EDL remains overcharged but the counterion accumulation near the surface reaches its saturation concentration determined by the excluded volume constraint. A three-dimensional condensed layer of counterions is formed with almost no coions. The thickness of the condensed layer increases as $\sigma$ becomes more negative. This phenomenon is commonly known as the ``crowding" of finite-size ions in EDLs\cite{Bazant2011DoubleCrowding, Kornyshev2007Double-layerChange,borukhovandelmansteric1997}. \par
\begin{table}
\centering
\begin{tabular}{ |c|c|c| }
\hline
\multirow{2}[3]{*}[2ex]{\textbf{Electrostatic Coupling}} & \multirow{2}[3]{*}[2ex]{\textbf{Dominant}} & \multirow{2}[3]{*}[2ex]{\textbf{Features of Electrical}} \\
\(\Xi = 2\pi q_\mathrm{+}^3l_\mathrm{b}^2\sigma/e\) & \textbf{Physics}& \textbf{ Double Layer} \\
\hline
Weak, $\Xi < 1$ & Mean-field effects as in PB & No overcharging \\
\hline
Moderate, $1 <\Xi < 100$ & Ion correlation & Overcharging increases with $\sigma$ \\
\hline
\multirow{2}{*}{Strong, $\Xi > 100$} & \multirow{2}[3]{*}[2ex]{Ion correlation +} & \multirow{2}[3]{*}[2ex]{Ionic Crowding/}  \\
& Excluded volume  & Overcharging reaches a plateau \\
\hline
\end{tabular}
  \caption{Electrical double layer behavior in three electrostatic coupling regimes}
  \label{tab:cregime}
\end{table}
To further elucidate the dependence of overcharging on $\sigma$, we define the degree of overcharging as $\sigma_\mathrm{ov} = \int_{0}^{z^*} \rho(z)dz   - |\sigma|$. $\sigma_\mathrm{ov}$ quantifies the number of accumulated counterions in excess of bare surface charge. $z^*$ is the position at which net charge density $\rho(z)$ changes sign from positive to negative. $\sigma_\mathrm{ov} = 0$ represents the absence of overcharging, whereas overcharging is stronger as $\sigma_\mathrm{ov}$ becomes more positive. Fig. \ref{fig:ov_sigma} shows a continuous transition from a normal double layer to an overcharged one, in agreement with the observations in experiments and simulations\cite{Diehl2006SmoluchowskiReversal, Diehl2008ColloidalConcentration, Martin-Molina2009EffectReversal}. At very high $\sigma$, $\sigma_\mathrm{ov}$ attains a plateau due to the saturation of counterion density near the surface. This trend has also been indicated by different independent simulation studies\cite{Diehl2006SmoluchowskiReversal,Tanaka2002ElectrophoresisStudy}, which observed a slow down of inverted $\psi_\mathrm{\zeta}$ and its potential saturation at high $\sigma$. The WKB approximation we used in our previous work, overestimates the strength of the ion
correlation and hence fails to capture the above moderate coupling regime and shows a discontinuous jump to the strong coupling regime (see Fig. 1 in Ref.\cite{agrawalciwkb}).\par
\begin{figure}
 \includegraphics[width=0.6\columnwidth]{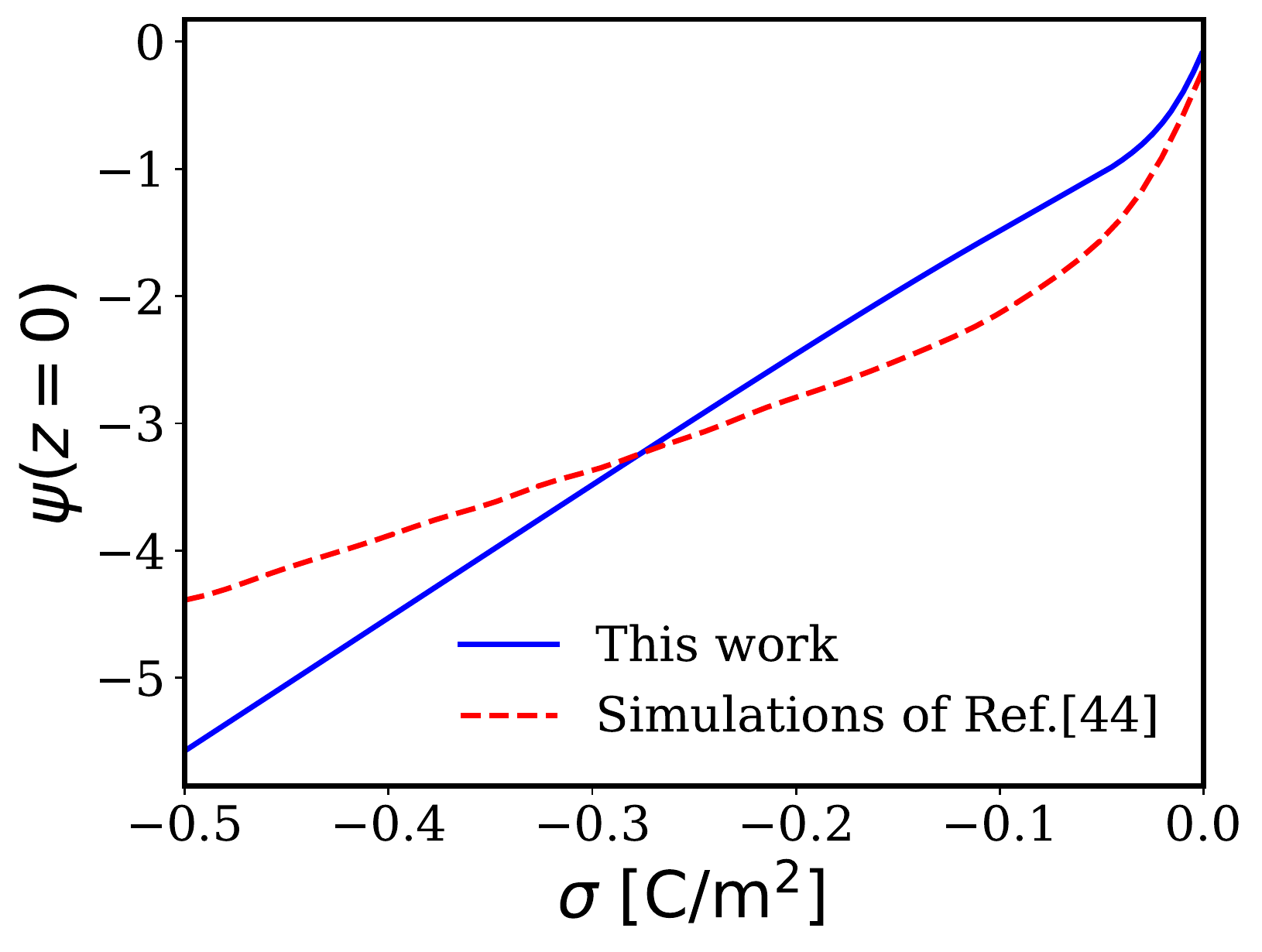}
  \caption{Electrostatic potential at the surface, $\psi(z=0)$, as a function of surface charge density $\sigma$. The solid lines represent predictions of our theory and the dashed lines represent the simulation results of Valisk\'{o} et al.\cite{valiskosim_2018}. Parameters used for both our calculations and simulations are: $c_\mathrm{b} = 0.1$ M, $a_\mathrm{\pm,s} = 1.5$ \AA, and $\varepsilon_\mathrm{S} = \varepsilon_\mathrm{P} = 80$.}
  \label{fig:psi_sim}
\end{figure}
The physical origin of overcharging can be characterized using the electrostatic coupling parameter $\Xi = 2\pi q_\mathrm{+}^3l_\mathrm{b}^2\sigma/e$, where $l_\mathrm{b}$ is the Bjerrum length. $\Xi$ quantifies the strength of correlations compared to thermal energy.
Fig. \ref{fig:ov_sigma} clearly shows that the overcharging curve can be divided into three regimes: $\sigma_\mathrm{ov} = 0$, a fast increase of $\sigma_\mathrm{ov}$, and the plateau, corresponding respectively to weak, moderate and strong coupling regimes. In the weak coupling regime, EDL can be qualitatively described by the mean-field PB. In the moderate coupling regime, PB fails to even qualitatively capture the overcharged EDL, which necessitates a systematic inclusion of correlations. Finally, in the strong-coupling regime, both correlations and excluded volume effects play a significant role in describing crowding. The EDL behavior in the three coupling regimes is summarised in Table \ref{tab:cregime}. Our theory predicts that the transition from weak to moderate coupling occurs at $\Xi \sim \mathrm{O}(1)$ and that from moderate to strong coupling occurs at $\Xi \sim \mathrm{O}(100)$. These results of transition points are in quantitative agreement with the values well-accepted in literature to separate different coupling regimes\cite{MoreiraStrong-couplingDistributions, Netz2001ElectrostatisticsTheory,najireviewwctosc}. Hence, our theory is successful in self-consistently unifying the description of overcharging in all three coupling regimes. We also note that a similar characterization of the overcharged EDL in terms of three distinct regimes was also done by Voukadinova and Gillespie \cite{voukadinova_gillespiedft}. However, to our knowledge, our theory is the first to discuss the existence of saturation in the degree of overcharging in the strong coupling regime. It is also important to highlight here that the inclusion of the excluded volume effect in the modified Gaussian renormalized fluctuation theory is essential to capture the crowding of finite-size ions and saturation in overcharging in the strong coupling limit.\par
\begin{figure}
 \includegraphics[width=0.6\columnwidth]{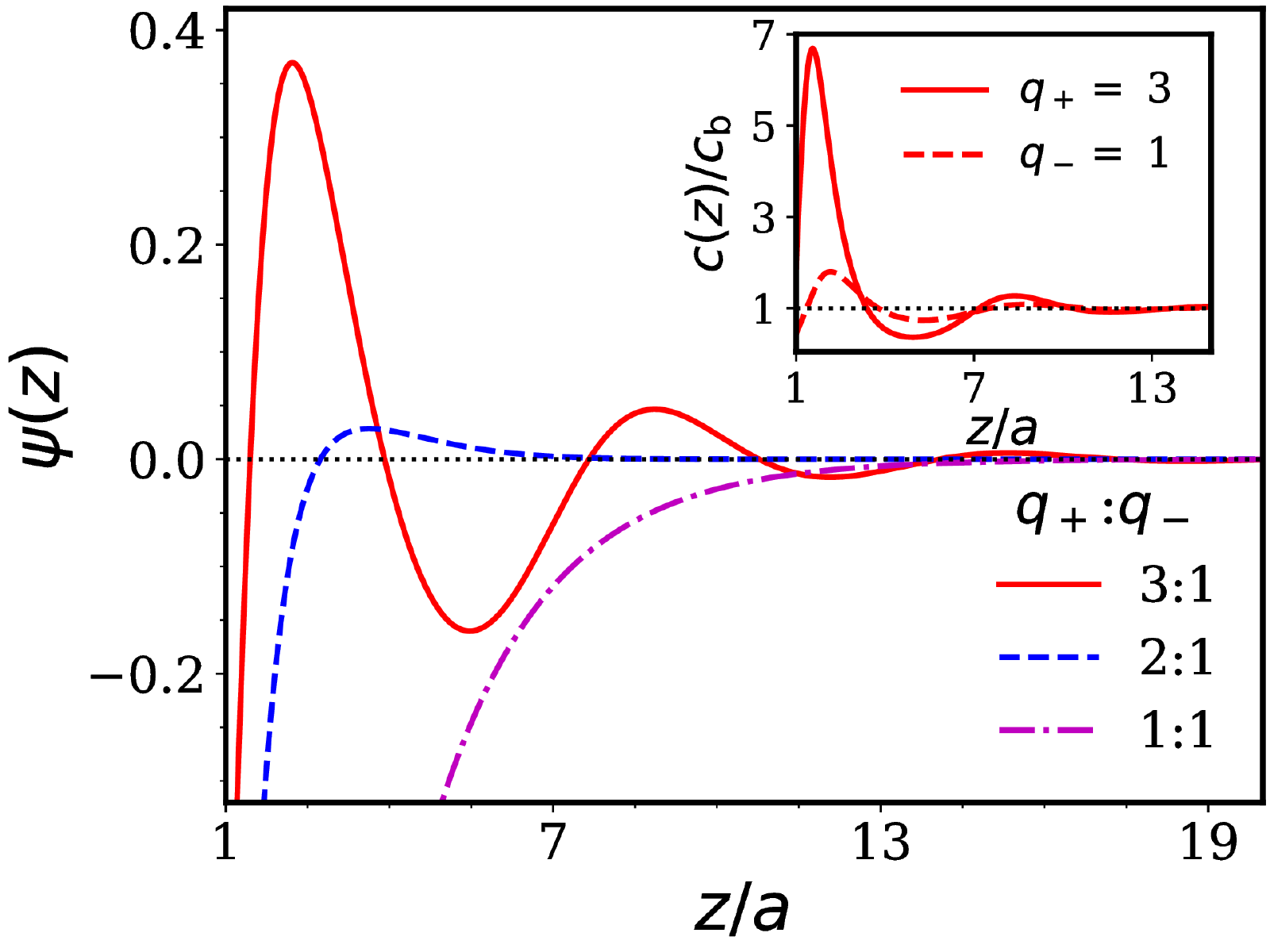}
  \caption{The effect of counterion valency on electrostatic potential profiles. The inset highlights the oscillations in counterion and coion distributions. $c_\mathrm{b} = 0.5$ M, $\sigma = - 0.15$ C/m$^2$, $a_\mathrm{\pm,s} = 2.1$ \AA, and $\varepsilon_\mathrm{S} = \varepsilon_\mathrm{P} = 80$.}
  \label{fig:psi_val}
\end{figure}

To validate our theory, we provide a quantitative comparison between our theory and Monte-Carlo simulations of Valisk\'{o} et al.\cite{valiskosim_2018} in Figure \ref{fig:psi_sim}. Using the same value of surface charge, ion size, and bulk salt concentration as in simulations, our theoretical predictions of surface electrostatic potential are in good agreement with the simulation data without any fitting parameters. The agreement is better at low $\sigma$ compared to high $\sigma$ values. One possible reason for this deviation could be the fact that the simulations of Valisk\'{o} et al. used an implicit model for solvent which ignores its excluded volume. This implicit treatment leads to larger counterion accumulation at the surface and hence lower surface potentials. We note that at high $\sigma$ the volume fraction of ions at the surface becomes very large. This requires a more rigorous treatment of the excluded volume effect compared to the local-density approximation used in our theory, such that the microstructure of the condensed layer can be quantitatively captured. However, in any case, the behavior of overcharging can still be divided into the three aforementioned coupling regimes as discussed above in Figure \ref{fig:ov_sigma}. In order to reproduce features like oscillations in counterion density profile as observed in simulations\cite{valiskosim_2018,dsouzabazantcrowding} the incompressibility constraint should be replaced by tools like the fundamental measure theory used in DFT-based approaches\cite{Gillespie2011EfficientlyTheory, gillespie2015review}.\par

Counterion valency also has a significant impact on overcharging. Increasing valency leads to stronger correlations, enhances overcharging, and even induces oscillations in electrostatic potential and ion distribution. In Fig. \ref{fig:psi_val}, $\psi(z)$ is plotted for $q_\mathrm{+} = $ 1, 2, and 3. Monovalent counterions do not show any overcharging even at very high $\sigma$. Although the correlation strength increases for divalent counterions, it results only in marginal overcharging for practical choices of surface charge and ion size. This is the reason why it is experimentally difficult to observe charge inversion in divalent electrolytes\cite{VanDerHeyden2006ChargeCurrents}. For trivalent ions, correlation is greatly enhanced, and overcharging is pronounced. Ionic layering occurs near the surface; oscillations in the distribution of both counterions and coions are observed as in the inset of Fig. \ref{fig:psi_val}. The oscillation is a sign of successive overcharging; each peak in the $\psi(z)$ essentially overcharges the layer of net negative charge preceding it. \par

Experiments\cite{VanDerHeyden2006ChargeCurrents,Martin-Molina2008ChargeSimulations} and simulations\cite{Hsiao2006Salt-inducedPolyelectrolytes,valiskosim_2018} have observed non-monotonic dependence of inverted ionic current and electrophoretic mobility on salt concentration, a feature which has not been fully understood yet. Here, we compare our theoretical predictions with experimental measurements of the streaming current $S_\mathrm{str}$ in planar nanochannels by van der Heyden et al.\cite{VanDerHeyden2006ChargeCurrents}. To calculate $S_\mathrm{str}$, we incorporate the double-layer structure into the Poiseuille flow. The expression for streaming current in a nanochannel is given by
 \begin{equation}
S_\mathrm{str} = w\int_0^H \rho(z)u(z)dz
\label{eq:s_str} 
 \end{equation}
 where $\rho(z)$ is the local charge density, $u(z)$ is the local fluid velocity, $w$ is the width and $H$ is the height of the nanochannel. An accurate theoretical prediction of streaming current requires knowledge of factors like the position of the slip plane, concentration-dependent viscosity, and ion correlations. Here we use a simple model with the position of the slip plane at $z=2a$, where $a$ is the radius of the ions. This choice for slip plane is commonly adopted in previous works\cite{Diehl2006SmoluchowskiReversal,Diehl2008ColloidalConcentration, Gillespie2011EfficientlyTheory} and is also supported by electrophoretic simulations \cite{Semenov2013ElectrophoreticSimulations, Lobaskin2007ElectrophoresisRegime}. The viscosity of the solution is taken to be that of the bulk water. For the case of pressure-driven flow, the local fluid velocity can be written from planar Poiseuille flow as 
 \begin{equation}
u(z) =\frac{\Delta P}{8\eta L}\frac{H^2}{(H/2-2a)^2}(z-2a)(H-2a-z)
\label{eq:velocity} 
 \end{equation}
 \begin{figure*}
\captionsetup[subfigure]{labelformat=empty}
    \begin{subfigure}{0.48\columnwidth}
      \includegraphics[width=\columnwidth]{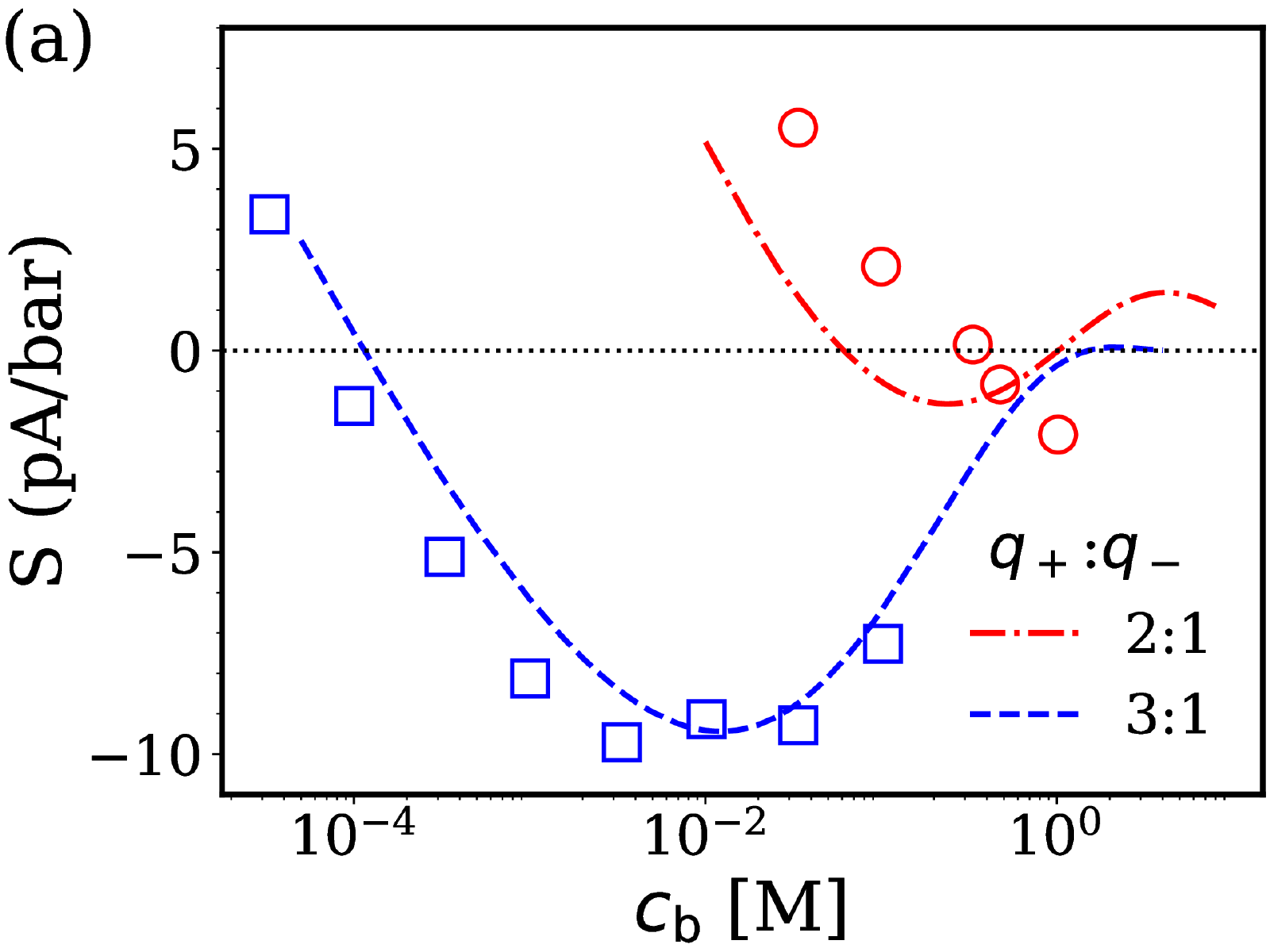} 
      \caption{}
      \label{fig:zeta_conc}
    \end{subfigure}
    \hfill
        \begin{subfigure}{0.48\columnwidth}
      \includegraphics[width=\columnwidth]{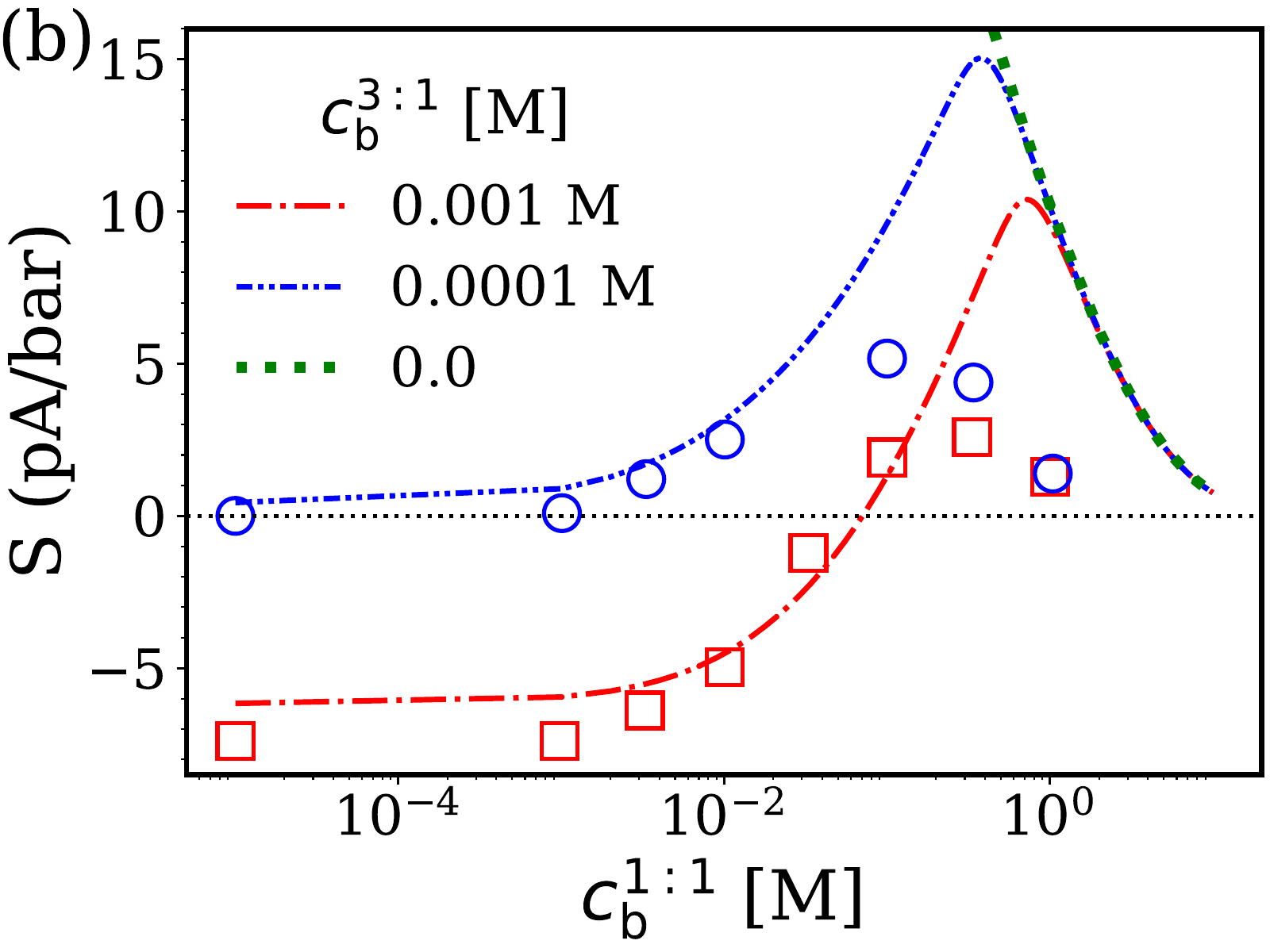}
      \caption{}
      \label{fig:zeta_mix}
    \end{subfigure}
\caption{Nonmonotonic dependence of charge inversion on salt concentration. $\sigma = -0.15$ C/m$^2$ and $\varepsilon_\mathrm{S} = \varepsilon_\mathrm{P} = 80$. a) Streaming current ${S}_\mathrm{str}$ as a function of $c_\mathrm{b}$ for pure divalent ($a_\pm = 1.5$ \AA) and trivalent ($a_\pm = 2.5$ \AA) salt solution. b) ${S}_\mathrm{str}$ as a function of added monovalent salt $c^\mathrm{1:1}_\mathrm{b}$ to a fixed trivalent salt concentration $c^\mathrm{3:1}_\mathrm{b}$. $a_\pm = 2.5$ \AA\ for both monovalent and trivalent salt. Lines represent our theoretical predictions and symbols represent ${S}_\mathrm{str}$ data adopted from van der Heyden et al.\cite{VanDerHeyden2006ChargeCurrents}.}
\label{fig:conc_panel}
\end{figure*}
where $\Delta P$ is the applied pressure difference, $\eta$ is the bulk viscosity of water and $L$ is the length of the nanochannel. Substituting Eq. \ref{eq:velocity} into Eq. \ref{eq:s_str} and using local charge density predicted by the theory, the streaming current can be calculated, as shown in Fig. \ref{fig:zeta_conc}. Only the ion size $a$ was used as a fitting parameter, and the surface charge and the dimensions of the nanochannel were adopted from the experimental setup of van der Heyden et al.\cite{VanDerHeyden2006ChargeCurrents}.\par

Fig. \ref{fig:zeta_conc} shows a non-monotonic behavior of $S_\mathrm{str}$ as a function of $c_\mathrm{b}$ for both divalent and trivalent salts. This nature is a consequence of competition between correlations and the translational entropy of ions. At low salt concentrations, the translational entropy loss for ions to accumulate at the surface is very large, which cannot be compensated by the energy gain from correlation. Thus, counterion accumulation is limited and there is no charge inversion. As c$_\mathrm{b}$ increases, the gain in correlation increases, whereas the entropic loss for ions to come to the surface decreases. As a result, counterion accumulation is sufficient to invert the sign of $\psi_\mathrm{\zeta}$ and hence $S_\mathrm{str}$ from positive to negative. For higher $c_\mathrm{b}$, the strength of ion correlations in bulk also increases which reduces the energetic incentive for the counterions to migrate to the surface. This together with the excluded volume effect at the surface leads to a maximum in $S_\mathrm{str}$. With the continued increase in $c_\mathrm{b}$, strong correlations in bulk further reduce counterion accumulation, and the sign of $S_\mathrm{str}$ changes back from positive to its original negative, manifested as ``reentrant charge inversion". Finally, at extremely high salt concentrations, $S_\mathrm{str}$ approaches zero due to strong screening. As shown in Fig. \ref{fig:zeta_conc}, our theoretical predictions capture the non-monotonic dependence of streaming currently on the salt concentration with a good quantitative agreement with experimental data \cite{VanDerHeyden2006ChargeCurrents}. The  $\psi_{\zeta}$ predicted by our theory for 3:1 salt is also close to the simulation results of Valisk\'{o} et al.\cite{valiskosim_2018}. The value of $\psi_{\zeta}$ at $c_\mathrm{b}$ = 1 M and $\sigma = -0.1$ C/m$^2$ as obtained in simulation is -0.06, and our theoretical result for the same $\sigma$, $c_\mathrm{b}$ and ion radius of 3.0 \AA\ is -0.024. This negative sign of $\psi_{\zeta}$ predicted by both simulations and our theory is consistent with the idea of non-monotonic behavior of charge inversion with salt concentration. For trivalent ions, the agreement is remarkable, because the electrostatic correlation is the dominant effect for ions with high valency. However, for divalent salts, our theory underestimates critical salt concentration for charge inversion. In the case of divalent counterions, the strength of correlations is not very strong and the structure of the double layer is also influenced by other effects like specific adsorption, the orientation of dipoles near the surface, and the hydration of ions.  \par

Our theory can also capture the non-monotonic reduction of charge inversion as monovalent salt is added to a multivalent salt solution\cite{VanDerHeyden2006ChargeCurrents}. The addition of monovalent salt also enhances ion correlations in bulk and hence in Fig. \ref{fig:zeta_mix}, $S_\mathrm{str}$ shows an initial increase and a subsequent disappearance of charge inversion as monovalent salt concentration $c^\mathrm{1:1}_\mathrm{b}$ increases. With the continued increase in $c^\mathrm{1:1}_\mathrm{b}$, the role of monovalent counterions becomes more important, resulting in a minimum in $S_\mathrm{str}$. After that, EDL gets dominated by monovalent salt, and $S_\mathrm{str}$ approaches zero. The curves of two $c^\mathrm{3:1}_\mathrm{b}$ values merge with that of a pure monovalent salt solution. At low $c^\mathrm{1:1}_\mathrm{b}$, our calculations of $S_\mathrm{str}$ are in excellent agreement with the experiments of Van der Heyden et al. \cite{VanDerHeyden2006ChargeCurrents}. The discrepancy between theory and experiments at high $c^\mathrm{mono}_\mathrm{b}$ could be attributed to the increased viscosity near the interface at high salt concentrations \cite{freundviscosity,qiaoaluru}. 
\begin{figure} 
 \includegraphics[width=0.6\columnwidth]{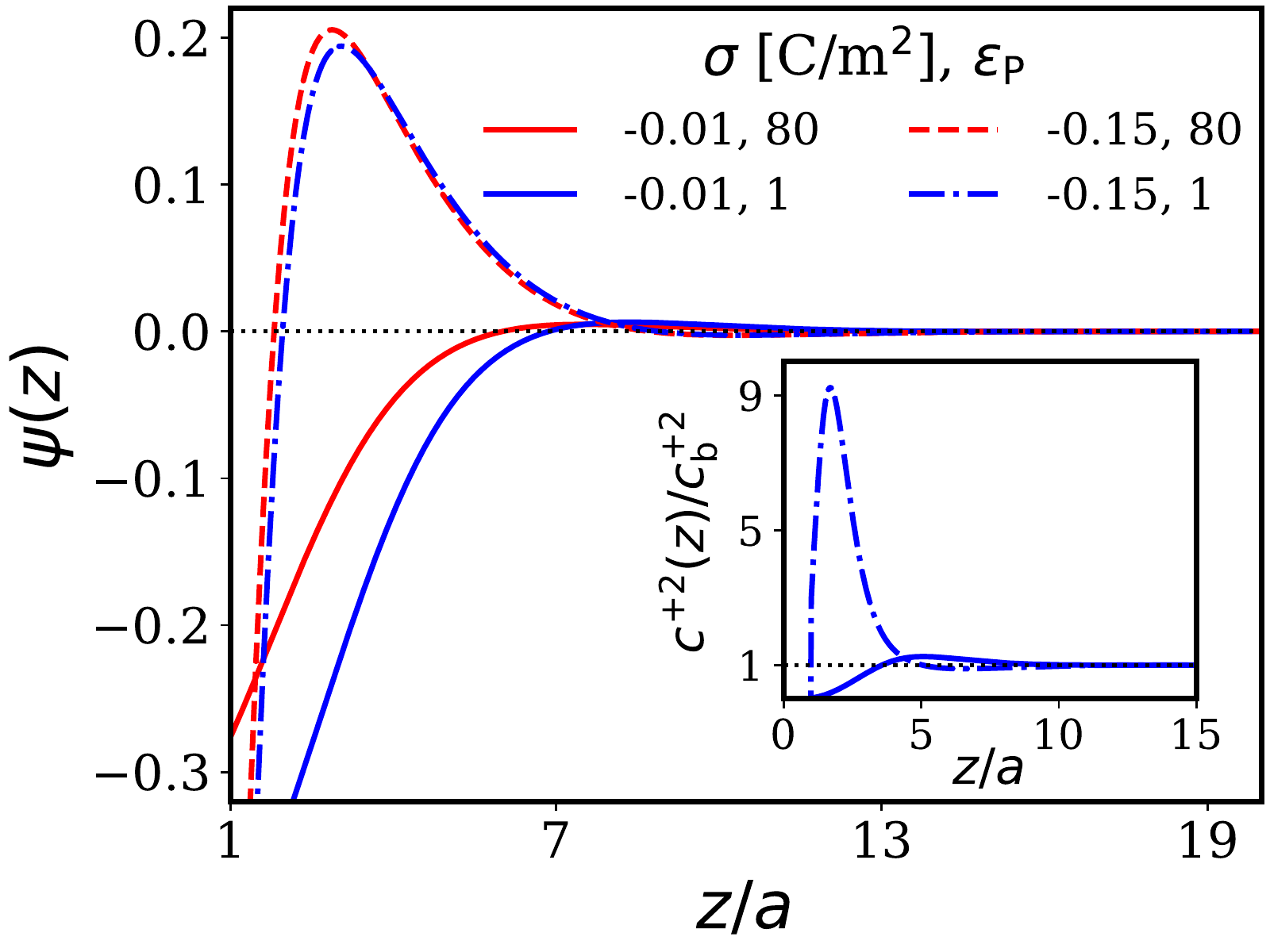}
  \caption{The image charge effect due to dielectric contrast on charge inversion. The inset shows the difference in counterion distribution under the absence ($\sigma = -0.01$ C/m$^2$) and the presence ($\sigma = -0.15$ C/m$^2$) of charge inversion when the dielectric contrast is included. $q_+ = 2$, $q_- = 1$, $c_\mathrm{b} =$ 0.5 M, $a_\mathrm{\pm,s} = 1.5$ \AA and $\epsilon_\mathrm{S} = 80 $.  }
  \label{fig:psi_image}
\end{figure}Issues with modeling electrokinetic flow at high salt concentrations are discussed in detail in the review paper by Bazant et al.\cite{bazantviscosity}. \par

The systematic treatment of electrostatic fluctuations allows us to simultaneously capture the inhomogeneity in both ionic strength and dielectric permittivity. In most real systems there is a dielectric mismatch between the charged plate and electrolyte solution, resulting in image charge repulsion on mobile ions. The image charge effect is found to alter EDL only in the weak coupling regime, refer to Fig. \ref{fig:psi_image}. {For 2:1 salt solution at low $\sigma = -0.01$ C/m$^2$, the electrostatic potential profile shifts significantly towards negative because the weak correlation due to low counterion concentration cannot counter the image charge depletion. On the contrary, for high $\sigma = -0.15$ C/m$^2$ when charge inversion occurs, the counterion concentration near the surface is so high that EDL is dominated by the ion correlation. The change in $\psi(z)$ is therefore almost negligible. These predictions are consistent with the simulation results of Wang and Ma \cite{Wang2010InsightsElectrolytes}. In Figure \ref{fig:image_panel} we compare our theoretical predictions with their results for the case of trivalent and monovalent salt mixture. At a high $\sigma$ of -0.16 C/m$^2$, the $\psi(z)$ profiles with the dielectric contrasts obtained from both theory and simulation completely overlap with the case of no dielectric contrast. \begin{figure*}
\captionsetup[subfigure]{labelformat=empty}
    \begin{subfigure}{0.48\columnwidth}
      \includegraphics[width=\columnwidth]{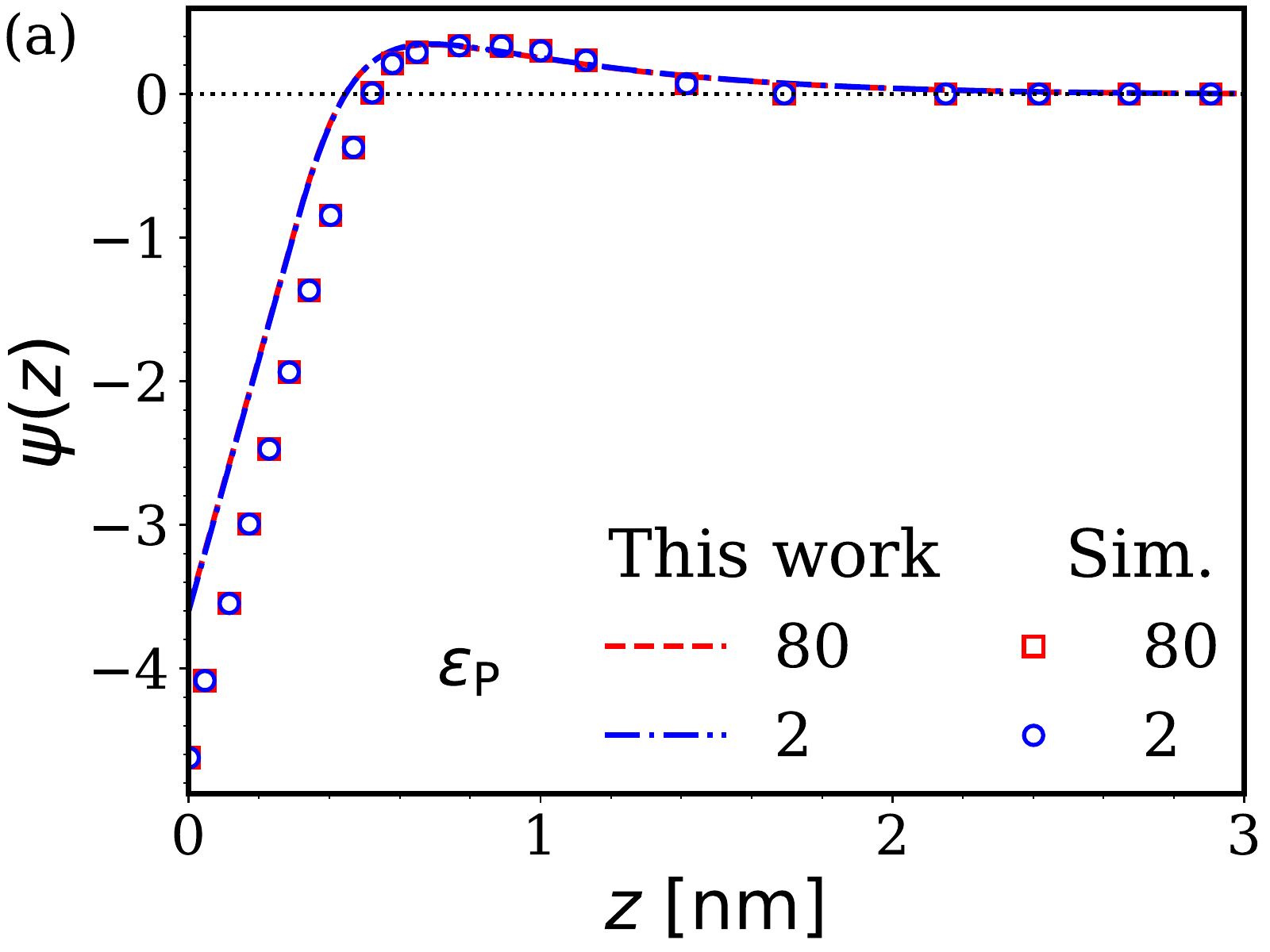} 
      \caption{}
      \label{fig:image_high}
    \end{subfigure}
    \hfill
        \begin{subfigure}{0.48\columnwidth}
      \includegraphics[width=\columnwidth]{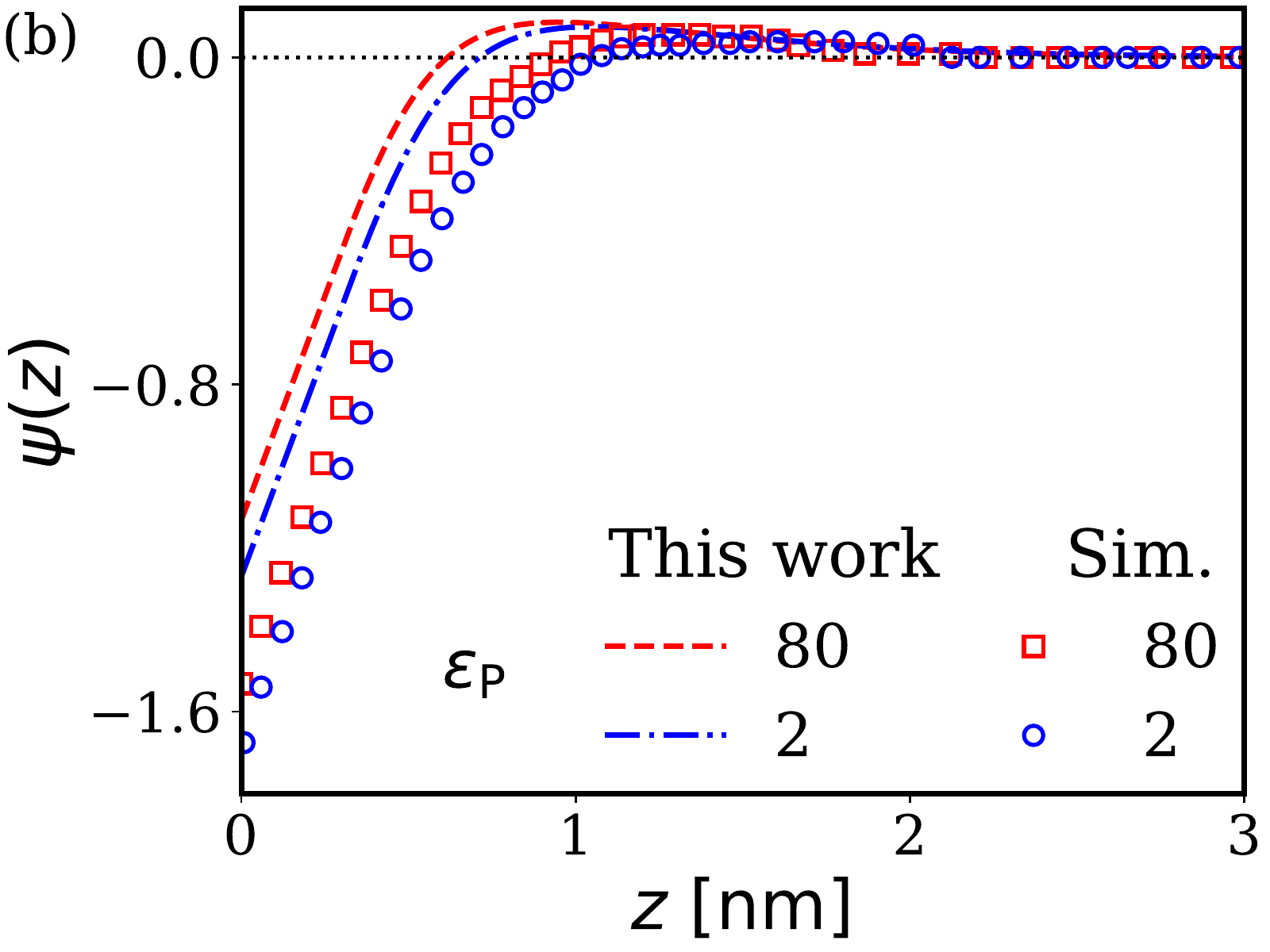}
      \caption{}
      \label{fig:image_low}
    \end{subfigure}
\caption{Effect of dielectric contrast on electrostatic potential profiles $\psi(z)$. Comparing our theoretical predictions with simulations of Wang and Ma\cite{Wang2010InsightsElectrolytes} for the case of trivalent and monovalent salt mixture. $c^\mathrm{3:1}_\mathrm{b} =$ 0.03 M, $c^\mathrm{1:1}_\mathrm{b} =$ 0.1 M, $a_\pm = 3.0$ \AA\ and $\epsilon_\mathrm{S} = 80 $. a)  $\sigma = -0.15$ C/m$^2$ and b) $\sigma = -0.04$ C/m$^2$. Lines represent our theoretical predictions and symbols represent simulation data.}
\label{fig:image_panel}
\end{figure*}However, at low $\sigma$ of -0.04 C/m$^2$, dielectric contrast was found to shift $\psi(z)$ in the negative direction. With only ion size $a_\pm = 3.0$ \AA\ as an adjustable parameter, the theoretical values of $\psi(z)$ in both cases are in good quantitative agreement with the simulations.\par

\section{Conclusion}
We have applied the modified Gaussian renormalized fluctuation theory to elucidate the nature of overcharging and charge inversion. The non-perturbative treatment of electrostatic fluctuation enables us to self-consistently capture the spatially varying ion correlation, dielectric permittivity, and excluded volume effect. Overcharging is dominated by ion correlations and excluded volume effects, with only a minor contribution from the image force. For multivalent electrolytes, increasing surface charge induces a continuous transition from a normal double layer to an overcharged one, and eventually to ionic crowding at the surface. These three characteristics of EDL correspond respectively to weak, moderate, and strong coupling regimes. Increasing counterion valency enhances overcharging and leads to ionic layering and oscillations. Our theory also correctly captures the non-monotonic dependence of charge inversion on salt concentration. The predictions of our theory are in good agreement with experimental and simulation results reported in the literature. Furthermore, being a field-theoretic formulation, our theory can be easily incorporated as the electrostatic component to study the structure and dynamic behaviors in a variety of soft matter, biophysical, and electrochemical systems. The ion correlation formulation developed here can be incorporated into the self-consistent field theory for polymers to model polyelectrolyte swelling and relaxation in multivalent salts\cite{Sing2014ElectrostaticMorphology,cduanprl}. Equation \ref{eq:mu_ions} provides a new expression for the electrochemical potential of ions in systems where mean-field PB is insufficient. This can enhance our understanding of ion solvation and transport in batteries and supercapacitors with multivalent ions\cite{wucharging2022}. Similarly, our theory can improve the potential of mean force calculations for ion permeation in biological ion channels. This could be particularly important for the transport of Ca$^{2+}$ and Mg$^{2+}$, where the electrostatic potential profile inside the channel induced by the surrounding baths cannot be accurately modeled using mean-field PB\cite{roux2004,floodchannelreview}. The capacity of our theory to correctly capture the salt concentration effect on ion correlations also makes it an ideal tool for modeling diffusiophoresis and diffusio-osmosis\cite{suinelectrokinetics2022}. Lastly, our theory can be employed to model the impact of spatially varying ion correlations on the structure of overlapping electric double layers and the associated free energies. This can provide insights into counterintuitive phenomena like opposite-charge repulsion and like-charge attraction in charged colloids\cite{agrawal2023lca}.
 
\begin{acknowledgement}

Acknowledgment is made to the donors of the American Chemical Society Petroleum Research Fund for partial support of this research. The authors also thank Prof. Carlo Carraro, Prof. Kranthi Mandadapu, and Dr. Dimitrios Fraggedakis of UC Berkeley for helpful discussions regarding the numerics of the Green's function and insightful comments on the results. This research used the computational resources provided by the Kenneth S. Pitzer Center for Theoretical Chemistry at UC Berkeley and the Savio computational cluster resource provided by the Berkeley Research Computing program.

\end{acknowledgement}

\begin{suppinfo}

See Supplemental Material for the derivation of the
modified Gaussian renormalized fluctuation theory, and the
method to decompose and compute the correlation function G.
\end{suppinfo}

\bibliography{ov_jpcb}

\end{document}